\documentclass[12pt]{article}
\usepackage{graphicx}
\large

\title{\large Masses of quantum particles as a result of diffeomorphisms}

\vspace{10 cm}

\author{ V.V. Belokurov$^{1,2}$ and E.T. Shavgulidze$^{1}$    \\
\\    {\small \em 1. Lomonosov Moscow State University, Russia }
\\    {\small \em 2. Institute for Nuclear Research of Russian Academy of Sciences, Russia }
\\ {\small  vvbelokurov@yandex.ru ; shavgulidze@bk.ru}}

\date{ \ \ \  }
\begin{document}
\maketitle

\begin{abstract}
We present the explicit form of the diffeomorphisms of the time variable that transform the measure for quantum free massless particle into the measure for quantum free massive particle .

\end{abstract}

\vspace{1cm}

The quasiinvariance of the Wiener measure\footnote{We consider Euclidean version of the theory and use Wiener integrals instead of Feynman path integrals.}
on the space of continuous functions 
\begin{equation}
   \label{M1}
 w(dy)=\exp\left\{ -\frac{1}{2}\int \limits _{a}^{b}\left(y'(t)\right)^{2}dt \right\}\ dy\,.
\end{equation}
under the diffeomorphisms of the interval was proven in the paper \cite{(Shepp)}. The Radon - Nikodym derivative was found there as well (see, also, \cite{(Shavgulidze1978)}).

Here we evaluate the Radon - Nikodym derivative in another way and 
 present the explicit form of the diffeomorphisms of the time variable that transform the measure for quantum free massless particle into the measure for quantum free massive particle .

Define the group of diffeomorphisms on the interval $[a,\,b]$
$$
g:\ [a,\,b]\rightarrow[a,\,b]\,;\ \ g(a)=a\,,\ g(b)=b\,.
$$
Let
$
g\in Diff^{3}_{+}\left([a,\,b] \right)\,,
$
that is, we suppose the function  $g(t)$  to be three times differentiable and $g'(t)>0\,.$

Note that $a$ and $b$ are arbitrary and  we can consider the group not only on the finite interval but on the axis $(-\infty,+\infty) $ or half-axis $(0, +\infty)$ as well.

The function  $y(\tau)$ is the image of the function $x(t)$  and is obtained according to the rule \cite{(Shepp)}, (see,also, \cite{(Neretin)})
\begin{equation}
   \label{M2}
y(\tau)=(gx)(\tau)=x(t)\sqrt{g'(t)}=x\,\left(g^{-1}(\tau)\right)
\,\frac{1}{\sqrt{\left(g^{-1}\right)'(\tau)}}\,.
\end{equation}

 After straightforward calculations\footnote{Since $x(t)$ is the Wiener process the integrals of the form $\int \limits _{a}^{b}
x(t)x'(t)f(t)\,dt$ are the Ito stochastic integrals (see, e.g. \cite{(McKean)}.} we get the equation
$$
\int\limits_{C\left([a,\,b]\right)} F[y]\,\exp \left\{-\frac{1}{2}\int \limits _{a}^{b}\left(y'(\tau)\right)^{2}\,d\tau\right\}\,dy=
$$
\begin{equation}
   \label{M3}
   C(g)\,
\int\limits_{C\left([a,\,b]\right)} F[gx]\,
\exp \left\{\frac{1}{4}\int \limits _{a}^{b}\,x^{2}(t)\mathcal{S}_{g}(t)\,dt+\frac{1}{4}\left(x^{2}(b)\frac{g''(b)}{g'(b)} -x^{2}(a)\frac{g''(a)}{g'(a)}\right)\right\}\,
\end{equation}
$$
\exp \left\{-\frac{1}{2}\int \limits _{a}^{b}\left(x'(t)\right)^{2}\,dt\right\}\,dx\,.
$$

Here $\mathcal{S}_{g}$ is the Schwarz derivative
$$
\mathcal{S}_{g}(t)\equiv
\left(\frac{g''(t)}{g'(t)}\right)'
-\frac{1}{2}\left(\frac{g''(t)}{g'(t)}\right)^2\,,
$$
 and the coefficient $C(g)$ depends on the diffeomorphism $g(t)$ and the type of the stochastic process $x(t)\,.$
 If the values $x(a)$ and $x(b)$ are fixed than
 \begin{equation}
   \label{M4}
 C(g)=\frac{1}{\sqrt{g'(a)g'(b)}}\,.
 \end{equation}
In other cases (unfixed one or both values) the coefficients $C(g)$ are different.

Now, for convenience, we suppose that $x(t)$ and $y(t)$ are Brownian bridges and $x(a)=x(b)=y(a)=y(b)=0\,.$

The diffeomorphisms of the form
\begin{equation}
   \label{M5}
g_{0}(t)=\frac{1}{\left(e^{2mb}-e^{2ma}\right)}\left\{(b-a)\,e^{2m t}+(a\,e^{2mb} - b \,e^{2ma} ) \right\}
\end{equation}
transform the interval to itself: $[a,\,b]\rightarrow [a,\,b]\,,$ and turn the Schwarz derivative into the constant
$$
\mathcal{S}_{g_{0}}(t)=-2\,m^{2}\,.
$$

The corresponding coefficient is
\begin{equation}
   \label{M6}
C(g_{0})=\frac{\sinh \left(m(b-a)\right)}{m(b-a)}\,.
\end{equation}

Thus, the equation  (\ref{M3})  looks like
$$
\int\limits_{C\left([a,\,b]\right)} F[y]\,\exp \left\{-\frac{1}{2}\int \limits _{a}^{b}\left(y'(t)\right)^{2}\,dt\right\}\,dy=
$$
\begin{equation}
   \label{M7}
\frac{\sinh \left(m(b-a)\right)}{m(b-a)}\,\int\limits_{C\left([a,\,b]\right)} F[gx]\,
\exp \left\{-\frac{1}{2}\int \limits _{a}^{b}\,m^{2}x^{2}(t)\,dt\ -\frac{1}{2}\int \limits _{a}^{b}\left(x'(t)\right)^{2}\,dt\right\}\,dx\,.
\end{equation}

This equation leads to the conclusion that seems paradoxical: \textbf{quantum theory of a free massive particle is nothing more than quantum theory of a free massless particle but with the different way of the measurement of time and the corresponding change of canonical coordinates}.

Due to the following property of Schwarz derivative
\begin{equation}
   \label{M8}
\mathcal{S}_{f\circ g}(t)= \mathcal{S}_{f}(g(t))\left(g'(t)\right)^{2}+ \mathcal{S}_{g}(t)\,,\ \ \ (\,f\circ g\,)(t)=f\left( g(t)\right)\,,
\end{equation}
there is a class of diffeomorphisms with $\mathcal{S}_{g}(t)=-2\,m^{2}\,,$ besides (\ref{M5}).

In fact, linear-fractional transformations
$$
f(t)=\frac{\alpha t+\beta}{t+\delta}
$$
have zero Schwarz derivative.

For $\alpha=(a+b)+\delta$ and $\beta=-ab $ they preserve the ends of the interval $[a,\,b]\,.$

Therefore, the general form of the diffeomorphisms leading to the equation  (\ref{M7}) is
\begin{equation}
   \label{M9}
g(t)=\frac{\left(a+b+\delta \right)g_{0}(t)-ab}{g_{0}(t)+\delta}
\end{equation}
where $g_{0}(t)$ is given by the equation (\ref{M5}) and $\delta$ is an arbitrary parameter.

Note that the coefficient (\ref{M6}) is invariant
$$
C(f\circ g_{0})=C(g_{0})\,.
$$

If we take $F=1$ in the equation (\ref{M7}) we get, as an additional bonus, the Feynman - Kac formula
without cumbersome evaluations of infinite products (see, e.g., \cite{(Simon)}).

The analog of the equation  (\ref{M7}) can be obtained if one makes the nonlocal but linear substitution in Wiener integrals
\begin{equation}
   \label{M10}
y(t)=x(t)+m\int \limits _{a}^{t}\,x(\tau)d\tau\,.
\end{equation}
In contrast to nonlocal and nonlinear substitutions  \cite{(BSh1)}, the substitution (\ref{M10}) does not lead the functional integration out of the space of continuous functions.

The substitutions in functional integrals accompanied by the change of the time variable make it possible to relate the quantum mechanical problems with the different potentials ( \cite{(Duru)}, \cite{(Inomata)}, \cite{(Ho)}, \cite{(Pak)} and others).
 In particular, in the paper \cite{(Storchak)} the relation between the quantum mechanical Green functions for the Hamiltonians obtained from each other by the transformations similar to diffeomorphisms was found.

 Now let us try to realise explicitly the above scheme in quantum field theory.
  In fact, the three-dimensional Fourier transform of the free field function represents the infinite set of quantum harmonic oscillators with the frequencies   $\sqrt{\vec{k}^{2}} $ and  $\sqrt{\vec{k}^{2}+m^{2}}\,, $ for the massless and the massive fields, respectively.
In this case, the transfer from massless relativistic particle to massive relativistic particle is realised by the composition of the diffeomorphisms
\begin{equation}
   \label{M11}
g_{\sqrt{\vec{k}^{2}+m^{2}}}\circ g_{\sqrt{\vec{k}^{2}}}^{-1}\,.
\end{equation}

Having in mind the infinite time interval we consider the diffeomorphism
\begin{equation}
   \label{M12}
g(t)=\exp\{2mt\}\,.
\end{equation}
Its Schwarz derivative equals to $-2m^{2}\,,$ and the inverse diffeomorphism has the form
\begin{equation}
   \label{M13}
g^{-1}(t)=\frac{1}{2m}\ln t\,.
\end{equation}

Now, the composition  (\ref{M11}) is
\begin{equation}
   \label{M14}
\tau _{k}=g_{\sqrt{\vec{k}^{2}+m^{2}}}\circ g_{\sqrt{\vec{k}^{2}}}^{-1}=g_{\sqrt{\vec{k}^{2}+m^{2}}}\left( g_{\sqrt{\vec{k}^{2}}}^{-1}(t)\right)=t^{\sigma (k)}
\end{equation}
where
$$
\sigma (k)=\sqrt{\frac{\vec{k}^{2}+m^{2}}{\vec{k}^{2}}}\,.
$$

Thus, for free scalar field we have
$$
\exp\{-\frac{1}{2}\int d^{3}\vec{k}\,\int dt\,\left(|\dot{\phi}(t,\vec{k})|^{2}+\vec{k}^{2}|\phi(t,\vec{k})|^{2} \right)\}\,d\phi=
$$
\begin{equation}
   \label{M15}
\exp\{-\frac{1}{2}\int d^{3}\vec{k}\ z(k)\,\int d\tau_{k}\,\left(|\dot{\varphi}(\tau_{k},\vec{k})|^{2}+\left(\vec{k}^{2}+m^{2}\right)|\varphi(\tau_{k},\vec{k})|^{2} \right)\}\,d\varphi\,.
\end{equation}
Here $ |\phi(t,\vec{k})|^{2}=\phi(t,\vec{k})\,\phi(t,-\vec{k}) $ and $z(k)$ is a normalizing factor.

However, the right-hand side of the equation
 (\ref{M15}) corresponds to a nonlocal theory. So, in quantum field theory we cannot relate to each other two similar theories (massive and massless) by a simple diffeomorphism. 

We are grateful to A.A. Slavnov and S.N. Storchak for valuable remarks.

\end{document}